\documentstyle[prb,aps,multicol]{revtex}
\begin{document}
\title{
Exact Finite-Size Spectra in the Kondo Problem \\
and Boundary Conformal Field Theory
}
\author{S. Fujimoto$\, ^1$, N. Kawakami$\, ^2$ and S.-K. Yang$\, ^3$}
\address{
$^1$Department of Physics, Faculty of Science,
Kyoto University, Kyoto 606, Japan\\
$^2$Department of Material and Life Science,
and Department of Applied Physics \\
Osaka University, Suita, Osaka 565, Japan\\
$^3$Institute of Physics, University of Tsukuba, Ibaraki 305, Japan
}
\maketitle

\begin{abstract}
The exact finite-size spectra for several quantum
impurity models related to the Kondo problem are obtained from
the Bethe ansatz solutions. Using the finite-size scaling
in boundary conformal field theory, we determine
various surface critical exponents from the exact spectrum,
which accord with those obtained by Affleck and
Ludwig with Kac-Moody fusion rules. Some applications to
critical phenomena observed in connection with the orthogonality catastrophe
are also discussed.
\end{abstract}


\vspace{1 cm}
\begin{multicols}{2}

\section{Introduction}

The method of two-dimensional conformal field theory (CFT) is very powerful to
study (1+1)-dimensional quantum critical phenomena\cite{bpz}.
One of important applications of CFT is due to
Affleck and Ludwig to analyse the Kondo effect.
As is well-known, the Kondo effect
has been providing many interesting issues
such as the multi-channel Kondo problem\cite{nb,tw,andrei}.
A new machinery developed in ref. 2 for studying the Kondo problem is
based on the Cardy's boundary CFT \cite{cardy}, and opened the way for
calculating correlation functions
and transport coefficients in the multi-channel Kondo model.
The CFT approach of Affleck and Ludwig, however,
is based upon a plausible hypothesis about the low-energy fixed point.
This is unavoidable since the formulation of quantum impurity problems in
terms of boundary CFT requires not only the macroscopic self-consistency
in CFT, but depends on some details of microscopic interactions.
They hence proposed the so-called fusion hypothesis
to describe the low-energy fixed point, from  which
the finite-size spectrum (FSS) was obtained to evaluate critical exponents.

On the other hand, the exact Bethe-ansatz solutions to various models
related to the Kondo problem have been known.
Thus it is desirable to  calculate the FSS analytically
and compare it with the results obtained by Affleck and Ludwig.
Motivated by this, we have recently computed
the FSS for the Anderson model, the {\it s-d}
exchange model, the SU($N$) Anderson model, and the multi-channel
Kondo model from the Bethe ansatz solution\cite{fky1,fk2}.
All the results obtained are in accordance
with those predicted by CFT.
Furthermore we can discuss various critical properties from
the exact spectrum using finite-size scaling
arguments\cite{aff1}.
In comparison with bulk systems, these quantum impurity models
show quite different critical behaviors reflecting the effect of
boundaries\cite{aff1,fky2}.

In this paper, we give a short review of our recent studies
on the exact spectrum for the Kondo problem.
The organization of the paper is as follows.
In Sec.II, we collect some basic results in boundary CFT
which will be needed in the following discussions.
In Sec.III, we obtain exactly the FSS of the Anderson model by
the Bethe ansatz method, and discuss its critical properties
characterizing the local Fermi liquid.
In Sec.IV, we show how these results
can be applied when considering
the Fermi-edge singularity in photoemission (absorption)
experiments. We then point out in Sec.V that such
anomalous exponents determine the correlation functions
of pseudo-particles in the Anderson model.
In Sec.VI, we derive a general expression for
the FSS of 1D multicomponent chiral systems, and discuss its
application to the X-ray edge singularity
in the edge state of the fractional quantum Hall effect.
In Sec.VII, we briefly discuss the FSS
of the multi-channel Kondo model
in the overscreening case where a non-Fermi
liquid fixed point plays a role. Finally a brief summary is given in Sec.VIII.

\section{Boundary Conformal Field Theory}

In this section, we summarize some basic
results in boundary CFT necessary for the following discussions.
For (1+1)-dimensional critical systems in
a finite geometry with open boundaries, we need to consider
conformal transformations which preserve the shape of the boundaries.
Then the anti-holomorphic part in CFT is not independent of
the holomorphic part, and critical properties are determined
only by the holomorphic part\cite{cardy}.

The critical behavior near the boundary is characterized by
the surface critical exponent $x_s$ which controls the
asymptotic behavior of the  correlation function,
$\langle \phi(t) \phi(0) \rangle \sim 1/t^{2x_s}$
($t\rightarrow \infty$). The surface  exponent
$x_s$ is generally different from bulk exponents.
We can regard the surface critical exponent $x_s$
as the conformal dimension of some boundary scaling operator.
Decomposing a scaling field $\phi$ into the holomorphic
part and the anti-holomorphic part which is defined by the
analytic continuation of the holomorphic part,
$\phi(z, \bar{z})=\phi_L(z)\phi_R(\bar{z})$,
we have $\phi(z, \bar{z})=
\phi_L(x+iy)\phi_L(x-iy)\sim y^{-2\Delta+\Delta_b}\phi_b(x)$.
Here $\Delta$ is the bulk conformal dimension of $\phi_L$, and
$\Delta_b$ is the conformal dimension of a boundary operator
$\phi_b(x)$. Then in the vicinity of the boundary, we have
the correlation function of the scaling operator,
\begin{equation}
\langle \phi (z, \bar{z})\phi(z', \bar{z}')\rangle \sim
\langle \phi_b(x)\phi_b(x')\rangle \sim \vert x-x'
\vert^{-2\Delta_b}.
\label{correlate}
\end{equation}
That is, the conformal dimension of the
boundary operator $\Delta_b$
directly determines the surface critical exponent $x_s$.

According to the finite-size scaling analysis in CFT \cite{cardy}
it is well-known that $\Delta_b$ enters in the finite-size corrections,
\begin{equation}
E=E_{\infty} -\frac{\pi vc}{24l}
+ \frac{\pi v}{l}(\Delta_b+n)
\label{eqn:fss}
\end{equation}
with $c$ being the central charge of the Virasoro algebra. Here
$E_{\infty}$ is the energy in the thermodynamic limit,
$l$ is the system size and $n$ is a non-negative integer
which features the conformal tower structure.

Since critical systems with boundaries involve only
the holomorphic part of CFT, we can describe the systems
only in terms of left-moving (or right-moving) currents.
Such systems are referred to as {\it  chiral systems}.
Since quantum impurity systems such as the Kondo model
can be mapped to 1D chiral systems, boundary CFT
is applicable \cite{aff1}.
In the rest of the paper, we shall
obtain the finite-size spectrum from the
exact solution of various models related to the Kondo problem,
and then discuss their critical properties using boundary CFT.

\section{Exact Finite-Size Spectrum in the Kondo problem}

There are several models related to the Kondo
problem, which have been exactly solved by the Bethe ansatz method.
So far thermodynamic quantities
have been mainly studied by the Bethe ansatz solution
\cite{bethe}, and the FSS has not been examined in detail.
As mentioned above, the FSS provides us with important
information on critical properties. Motivated by
this, applying standard methods in the Bethe ansatz solution,
we have systematically calculated the FSS
of the Anderson model, the SU($N$) Anderson
model and the {\it s-d} exchange model, and compared
the obtained results with the predictions by CFT\cite{fky1}.
In this section, we summarize  the results for the single-impurity
Anderson model as an example, since the manipulations
in deriving the FSS are essentially the same for these models.
The Hamiltonian of the Anderson model is given by
\begin{eqnarray}
H&=&\sum_{k,\sigma}\epsilon_{k}c_{k\sigma}^{\dagger}c_{k\sigma}
+V\sum_{k,\sigma}(c_{k\sigma}^{\dagger}d_{\sigma}+
d_{\sigma}^{\dagger}
c_{k\sigma}) \nonumber \\
&&+\epsilon_{d}
\sum_{\sigma}d^{\dagger}_{\sigma}d_{\sigma}
+Ud^{\dagger}_{\uparrow}d_{\uparrow}
d^{\dagger}_{\downarrow}d_{\downarrow},
\label{eqn:haman}
\end{eqnarray}
with standard notations \cite{tw}. The model describes
free conduction electrons  coupled with
correlated $d$-electrons at the impurity site via the resonant
hybridization $V$. After reducing the model to the
one-dimensional system with the use of
partial wave representation, we  linearize the
spectrum of  conduction electrons as $\epsilon_k =vk$
near the Fermi point.  The kinetic energy is
then replaced by the operator $-i v \partial /\partial x$
in the coordinate representation.
These simplifications enable us to apply  the Bethe-ansatz
method to diagonalize the Hamiltonian \cite{bethe,wieg,kawa}.

\subsection{Finite-size spectrum}

Exploiting standard techniques in Bethe-ansatz
method\cite{finite}, we can compute the finite-size corrections
to the ground-state energy as,
\begin{equation}
E_0=L \varepsilon_0-\frac{\pi v_c}{12L}-\frac{\pi v_s}{12L}~,
\label{eqn:andcasimir}
\end{equation}
where $L \varepsilon_0$ is the ground state energy
in the thermodynamic limit ($L\rightarrow \infty$), and
$v_c$ and $v_s$ are the velocities of massless charge
and spin excitations. For the present model,
these velocities take the same value $v$.
To compare the result (\ref{eqn:andcasimir}) with
the finite-size scaling formula
in CFT \cite{cardy}, one has to
replace $L$ by $2l$ since $L$ has been defined as the
periodic length of the system. Then we indeed find the scaling
behavior predicted by  boundary CFT \cite{cardy}
for the charge and spin sectors in (\ref{eqn:andcasimir}),
i.e. their  Virasoro central charges are given by $c=1$.

Let us now consider the excitation energy.
In the present system, there exist particle-hole type
excitations as well as  excitations which change the number
of electrons $N_h$ as well as  the $z$-component of spin $S_h$,
\begin{equation}
N_h \rightarrow N_h +\Delta N_h, \qquad S_h \rightarrow S_h+\Delta S_h.
\end{equation}
All these excitations give $1/L$-corrections to the energy spectrum.
An important point in the Kondo problem is
that the first-order corrections to the energy due to
the change of $N_h$ appear in $1/L$-corrections as an
 effect due to the impurity. This correction is given by
\begin{equation}
\Delta E^{(1)}=-\frac{2\delta_F}{L}\Delta N_{h},
\label{eqn:chemical}
\end{equation}
where $\delta_F$ is the phase shift caused by the
impurity scattering.
Combining this term with the  second-order corrections in $\Delta N_h$
and $\Delta S_h$, we obtain the excitation spectrum,
\begin{equation}
E=E_0+\frac{1}{L}E_1+\frac{1}{L^2}E_2+O(1/L^3),
\label{eqn:scaling}
\end{equation}
\begin{eqnarray}
\frac{1}{L}E_1&=&\frac{2\pi v}{L}
\biggl[\frac{1}{4}\biggl(\Delta N_{h}-2\frac{\delta_F}{\pi}
\biggr)^2
+n^{+}_c\biggr] \nonumber \\
&&+\frac{2\pi v}{L}[(\Delta S_{h})^2+n^{+}_s],
\label{eqn:finene1}
\end{eqnarray}
\begin{eqnarray}
\frac{1}{L^2}E_2&=&\frac{2\pi v}{L^2}
\frac{\chi_{c}^{imp}}{\chi_{c}^{h}}
 \biggl[\frac{(\Delta N_h)^2}{4}+n^{+}_c\biggr] \nonumber \\
&&+\frac{2\pi v}{L^2} \frac{\chi_{s}^{imp}}{\chi_{s}^{h}}
[(\Delta S_h)^2+n^{+}_s],
\label{eqn:finene2}
\end{eqnarray}
where $\chi_c^{imp}$ ($\chi_s^{imp}$) and $\chi_c^h$ ($\chi_s^h$)
are the charge (spin) susceptibilities of
impurity and host electrons, respectively.
$n^{+}_{c,s}$ are non-negative integers
which label particle-hole excitations. One can see that
the expression for the $1/L$ correction is
consistent with the fusion hypothesis\cite{aff1}.
Using the above formulas for the FSS, we will discuss various
critical properties of the Kondo model in the following.

\subsection{Canonical exponents}

Let us now  discuss the  results
(\ref{eqn:scaling})$\sim$(\ref{eqn:finene2}) by using
the finite-size scaling in CFT. It was already
found that low-energy critical properties
of the Kondo effect can be described by boundary
CFT in which we have  only the right (or left) moving sector of
CFT \cite{aff1}. The $1/L$ correction term
(\ref{eqn:finene1}) indeed shows the scaling
behavior predicted by boundary CFT.
It is seen from eqs.(\ref{eqn:andcasimir})
and (\ref{eqn:finene1}) that
the charge sector is described by $c=1$  Gaussian CFT
and the spin sector by
$c=1$ SU(2) Kac-Moody CFT in the low-energy regime.
Note that the finite-size spectrum for bulk
electrons in eq.(\ref{eqn:finene1}) involves a
non-universal phase shift $\delta_F$. We recall here
that this phase shift $\delta_F$ is regarded as
the chemical-potential change due to the
impurity (see eq.(\ref{eqn:chemical})).
This means that the effect of the phase
shift amounts to merely imposing twisted boundary conditions
on conduction electrons \cite{aff1}.
Therefore, when we derive the dimension
of the scaling operator associated with conduction electrons, we
should discard  the $\delta_F$ dependence
in eq.(\ref{eqn:finene1})
by redefining $\Delta N_h-2\delta_F/\pi\rightarrow\Delta N_h$.
Hence the scaling dimension $x$ of the conduction electron field
is obtained by taking the quantum numbers
\begin{equation}
\Delta N_h=1, \hskip 5mm \Delta S_h= 1/2,
\label{eqn:quanta}
\end{equation}
resulting in $x=1/2$.
Thus we can see that the single-electron Green function
$\langle c_{\sigma}(t)c^{\dagger}_{\sigma}(0)\rangle\sim 1/t^{\eta}$
has the canonical exponent $\eta=1$.
This result is consistent with the fact that the
system is described by the strong-coupling fixed point of
the local Fermi liquid \cite{fermi,aff1}.

\subsection{Local Fermi liquid}

The local Fermi-liquid properties are confirmed further
by observing the  $1/L^2$ corrections in eq.(\ref{eqn:finene2}).
For this, let us rewrite the $1/L^2$-term in eq.(\ref{eqn:finene2})
by using the velocities for the charge and spin excitations
at the impurity site,
\begin{equation}
v_c^{imp}=\frac{\pi }{3\gamma_c^{imp}},
\hskip 5mm
v_s^{imp}=\frac{\pi }{3\gamma_s^{imp}}.
\end{equation}
We then have
\begin{equation}
E_2 = \pi v_c^{imp}\frac{(\Delta n_d)^2}{4}
+2\pi v_{s}^{imp}(\Delta s_z)^2,
\label{eqn:fscimp}
\end{equation}
where we set  $n^{+}_c=n^{+}_s=0$ for simplicity.
Here $n_d$ is the number of impurity electrons, and
$s_z$ is the $z$-component of the impurity spin.
It should be noted that
this expression for the impurity spectrum
directly reflects  local Fermi-liquid properties.
To see this, we first recall that
in the ordinary Tomonaga-Luttinger liquid in 1D, there
is a dimensionless coupling parameter $K_{\rho}$,
which enters in the FSS as
$v_c K_\rho$, in addition to the velocities $v_c$ and $v_s$.
In the expression (\ref{eqn:fscimp}), however,
it is seen that $K_{\rho}=1$, {\it i.e.} it is given by
the value of free electrons. Moreover, $K_{\rho}=1$
always holds independent of
the strength of interaction $U$, which
suggests that the local Fermi liquid is stable
at the impurity site\cite{fermi,yamada}.
Although the spin-charge separation, which is characterized by the
different spin and charge velocities,
occurs even in the Kondo system,
the local Fermi liquid is always stabilized due to
the locality of correlations.

\section{Anomalous Exponents Related to the X-ray Edge Problem}

In the previous section, we neglected the effect of the
phase shift when  obtaining  the
canonical exponents for the local Fermi liquid. Keeping the phase-shift
dependence, on the other hand,
we can extract another interesting
information from the finite-size spectrum
(\ref{eqn:finene1}),  i.e. the critical behavior
related to the orthogonality catastrophe.
To see this explicitly, let us consider the
time-dependent Anderson model in which
the hybridization $V=0$
for $t<t_0$, and then $V$ is switched on at $t=t_0$.
Note that the orthogonality catastrophe related
to the Fermi edge singularity appears in
the long-time behavior of this model.
In this case, the phase shift in
(\ref{eqn:finene1}) becomes  a key quantity
which controls the critical exponent. Actually, critical
exponents related to the X-ray problem can be read off from
the above FSS with keeping the dependence on the
phase shift intact \cite{fky1,afflud}.
This is because a sudden potential change occurs
in X-ray photoemission (or absorption) experiments.
For example, the critical exponent $\eta$ of
the single-particle Green function
$\langle c_{\sigma}(t)c_{\sigma}^{\dagger}(0)\rangle
\sim 1/t^{\eta}$ $(0<t_0 \ll t)$
for the  model with a sudden potential change
is obbained as \cite{afflud}
\begin{equation}
\eta=1-\frac{2\delta_F}{\pi}+2\biggl(\frac{\delta_F}{\pi}\biggr)^2.
\end{equation}
Note that this is just the exponent which governs the
long-time behavior of the overlap integral  between
the initial and final states in
the X-ray absorption problem.
It is thus seen that the FSS (\ref{eqn:finene1})
contains the information about orthogonality catastrophe
in addition to the local Fermi-liquid properties.
In the next section, we shall discuss in more detail
the physical meaning of such anomalous exponents
in terms of pseudo-particles in the Anderson model.

\section{Critical Exponents of Pseudo-Particles in the Kondo Problem}

In the strong-coupling limit ($U\rightarrow\infty$)
of the Anderson model, the double occupancy of electrons
is forbidden, and thus the Fock space of the impurity electron
can be mapped to that spanned by the slave-boson field $b$
and the pseudo-fermion field $f_m$
($m=1,2, \cdots, N$) which represent an empty site
and a singly occupied site, respectively. In this section,
we discuss the long-time asymptotic behavior of dynamical
correlation functions of these pseudo-particles, and show
that the orthogonality catastrophe mentioned in the previous section
manifests itself in these correlation functions. The
critical exponents of pseudo-particles can be thus derived
exactly from a boundary CFT analysis.

Let us consider the $U \rightarrow \infty$ SU($N$) Anderson model.
The Hamiltonian is given by
\begin{eqnarray}
H & = &\sum_{m=1}^N \int{\rm d}x c^{\dagger}_{m}(x)\Big( -i
\frac{\partial}{\partial x} \Big) c_{m}(x)
+\epsilon_{f}\sum_{m=1}^N f_m^{\dag}f_m    \nonumber \\
&+& V\sum_{m=1}^N \int{\rm d}x\delta (x)
\Big( f_m^{\dag} b c_{m}(x)+c_{m}^{\dagger}(x)
b^{\dag} f_m \Big),
\label{eqn:hamsun}
\end{eqnarray}
with the constraint $b^{\dagger}b+\sum_{m}f^{\dagger}_{m}f_{m}=1$,
where the impurity electrons have $N$-fold spin states.
The Hamiltonian (\ref{eqn:hamsun}) can be diagonalized by
the Bethe-ansatz method after reducing it to
the one-dimensional one as mentioned before\cite{sch}.
The finite-size spectrum is then computed by standard
techniques\cite{finite}. The result is succinctly expressed in terms of
the $N \times N$ matrix\cite{fky1},
\begin{equation}
\frac{1}{L}E_1=
\frac{2\pi v}{L}\frac{1}{2}\Delta {\bf M}^{T}
{\cal C}_{f}\Delta {\bf M}
-\frac{\pi v}{L}N
\biggl(\frac{\delta_F}{\pi}\biggr)^2, \label{eqn:fss1}
\end{equation}
where  $\Delta M^{(l)} \equiv
\Delta M_h^{(l)}-\frac{\delta_F}{\pi}(N-l)$ for $1\leq l\leq N-1$,
and $\Delta  M^{(0)}=\Delta N_h-N\delta_F/\pi$
with $\delta_F$ being the phase shift at the Fermi level.
Here $\Delta N_h$ is the quantum number for charge excitations,
whereas $\Delta M_h^{(l)}$'s are quantum numbers
for spin excitations. The $N \times N$ matrix
${\cal C}_{f}$ is given as
\begin{equation}
{\cal C}_{f}=
\left(
\matrix {1     & -1      & \null  &
                 \smash{\lower1.7ex\hbox{\LARGE 0}} \cr
        -1     &  2  & \ddots & \null   \cr
        \null  & \ddots  &  \ddots     & -1      \cr
        \smash{\hbox{\LARGE 0}}   & \null   &  -1    & 2  \cr}
\right) .
\label{eqn:cf}
\end{equation}
It  can be checked that the last term in eq.(\ref{eqn:fss1}),
which is evaluated from the excited states,
is equal to the shift of the ground-state energy due to the presence
of the impurity. Therefore the increase
of the ground-state energy exactly
cancels the last term of eq.(\ref{eqn:fss1}),
which is not necessary for
the following discussions of critical exponents.
We will drop it in the following.

Let us now study the long-time behavior of the
Green functions for pseudo-particles;
$\langle f^{\dagger}_{m}(t)f_{m}(0)\rangle\sim t^{-\alpha_f}$, and
$\langle b^{\dagger}(t)b(0)\rangle\sim t^{-\alpha_b}$.
As explained in Sec. III, when determining canonical exponents
for the local Fermi liquid,
we can neglect the phase shift in eq.(\ref{eqn:fss1}).
In order to derive critical exponents for pseudo-particles,
however, we must regard the number of impurity electrons
(or phase shift) $n_l=\delta_l/\pi$ as  a quantum number.
Therefore the phase shift plays an essential role to determine the
critical exponents.  For example, in order to obtain
the Green function of pseudo-fermions,
we take  $\Delta N_h=1$ and
$\Delta M_h^{(l)}=0$ as quantum numbers. We thus obtain
the corresponding critical exponent as,
\begin{equation}
\alpha_f=1-\frac{2\delta_F}{\pi}+N\biggl(\frac{\delta_F}{\pi}
\biggr)^2.
\label{eqn:xray1}
\end{equation}
In a similar way, the critical exponent $\alpha_b$ for the
slave-boson Green function
can be obtained. Since the
slave-boson expresses a vacancy, it carries neither charge nor spin.
Putting  $\Delta N_h=\Delta M_h^{(l)}=0$, one gets
\begin{equation}
\alpha_b=N\biggl(\frac{\delta_F}{\pi}\biggr)^2.
\label{eqn:xray2}
\end{equation}
These expressions
for $\alpha_f$ and $\alpha_b$ agree with those obtained
for the $N=1$ and $N=2$ cases\cite{menge,costi},
and take the same form as those in the X-ray problem:
the exponent of pseudo-fermion corresponds to the X-ray
absorption exponent, and that of slave-boson
to the X-ray photoemission exponent.
In the Kondo effect, the Fermi edge singularity shows up
in the intermediate state as pointed out  in the
Anderson-Yuval approach. The anomalous exponents discussed here
reflect this singularity.

\section{Surface Critical Exponents in Chiral 1D Systems}

The expression for the FSS obtained for the Kondo problem
is essentially common to all 1D chiral quantum systems\cite{fky2}.
A number of 1D quantum systems with open boundaries
belong to this class.
For instance, let us list several examples of
exactly solvable models with open boundaries;
the Heisenberg model\cite{bxxz},
the boson model\cite{gaud}, the continuum electron model\cite{woy},
the Hubbard model\cite{shu}, and the
$1/r^2$ quantum model\cite{yam}, etc.
In contrast to the Kondo problem, however, there is a different
point in these 1D critical systems with boundaries.
Namely, in 1D  quantum models, a continuously
varying parameter $K_{\rho}$ enters in the theory,
reflecting U(1) symmetry of the charge sector. For
example, the critical exponent of the SU($N$) interacting
electron model with open boundaries is given by\cite{fky2,fk1}
\begin{equation}
x_b=\Delta {\bf M}^{T}
\left(
\matrix {\frac{1}{NK_{\rho}}+\frac{N-1}{N}     & -1      & \null  &
                 \smash{\lower1.7ex\hbox{\LARGE 0}} \cr
        -1     & 2  & \ddots & \null   \cr
        \null  & \ddots  &  \ddots     & -1      \cr
     \smash{\hbox{\LARGE 0}}   & \null   &  -1    & 2  \cr}
\right)
\Delta {\bf M},
\label{exponent}
\end{equation}
{}from which one can see, by comparing it with
(\ref{eqn:cf}),  that the only the charge
sector of the matrix is changed. Here the phase shift
$\delta_l$ implicitly involved in $\Delta M^{(l)}$
(see eq. (\ref{eqn:fss1})) can be
determined by the number of particles with spin $l$
localized at boundaries, $n_l=\delta_l/\pi$. From this
spectrum, we can read critical exponents of 1D
quantum systems with open boundary conditions.
The formula (\ref{exponent})
is typical for 1D {\it chiral} electron systems
with SU($N$) spin symmetry.

Following the arguments given in the previous section,
we can apply the above results to the X-ray problem in
1D chiral electron systems.
Suppose that electrons move only in one direction and
the backward scattering due to the impurity is irrelevant.
Such a situation may be realized
in the edge state of the fractional quantum Hall
effect(FQHE)\cite{wen}.
Let us discuss the X-ray photoemission (or absorption)
 problem in such chiral systems by applying the
formula (\ref{exponent}). For example, the critical
exponent for the X-ray absorption is given by\cite{fky2,fk1},
\begin{equation}
\alpha_f=\frac{1}{NK_{\rho}}\biggl(1-\frac{N\delta}{\pi}\biggr)^2
+\frac{N-1}{N},
\end{equation}
whereas the exponent for the photoemission is
\begin{equation}
\alpha_b=\frac{N}{K_{\rho}}\biggl(\frac{\delta}{\pi}\biggr)^2.
\end{equation}
These expressions should be compared with
(\ref{eqn:xray1}) and (\ref{eqn:xray2}).
Note that for the edge state of the FQHE
with filling $\nu=N/(Nm+1)$ with even $m$,
$K_{\rho}$ is solely determined by the  filling factor $\nu$ as
$K_{\rho}=\nu/N$. We expect such anomalous exponents to be
observed in the X-ray problem
in the edge state of the FQHE.

\section{Exact Finite-Size Spectrum of the Multi-Channel Kondo
Model}

We have seen so far that the FSS for the
ordinary Kondo effect is in complete accordance with
the fusion hypothesis. In this section, we extend our discussions to
the multi-channel Kondo model\cite{fk2}.
The Hamiltonian for the multi-channel model is
\begin{equation}
H = \sum_{k, l, \sigma} \epsilon_k c_{kl\sigma}^{\dagger}
 c_{kl \sigma}
    + J \sum_{k,k',l,\sigma,\sigma '} c_{kl\sigma}^{\dagger}
 (\sigma_
      {\sigma \sigma '} \cdot {\bf S}) c_{k'l\sigma '}
\end{equation}
with $J>0$,
where conduction electrons with  $n$-channels ($l=1, 2, \cdots, n$)
screen the impurity spin $S$.
Although the multi-channel model is
Bethe-ansatz solvable, it is quite difficult to
calculate  the FSS analytically, in particular, for the
overscreening case $n>2S$. The difficulty arises from the fact that
the Bethe-ansatz solution to the  multi-channel model takes the form of
the so-called string solution
even for the ground state. It has been known that the string solution is
valid only in the
thermodynamic limit\cite{tw,andrei}. Thus,
a naive application of finite-size techniques
\cite{finite} fails, giving only the Gaussian part of the
spectrum  for the spin sector. Namely,
the total spectrum has a pathological form,
\begin{eqnarray}
E=\frac{2\pi v}{L} & \biggl( & \frac{(Q-n)^2}{4n}+\frac{S_z^2}{n}
+(\mbox{flavor part}) \nonumber \\
&&+n_Q+n_s+n_f\biggr),\label{eqn:mulfss1}
\end{eqnarray}
for the finite system with linear size $L$,
where  $v$ is the Fermi velocity, and
$n_Q$, $n_s$, and $n_f$ are non-negative integers.
Here the first term expresses charge excitations
with the quantum number $Q$, whereas
the second term labels the spin excitations
with the magnetization $S_z$.
Since the spin sector of the multi-channel Kondo model
is described by the level-$n$ SU(2) Kac Moody theory
with the central charge $c_{WZW}=3n/(2+n)$ \cite{aff1},
it is seen from (\ref{eqn:mulfss1})
that the $Z_n$ parafermion sector with $c=c_{WZW}-1=2(n-1)/(n+2)$
is missing. It is thus necessary to
exploit alternative methods other than the
coordinate Bethe ansatz to obtain the correct spectrum corresponding to
the missing Z$_n$ parafermions.

We propose an analytic way to
investigate the nontrivial Z$_n$ parafermion part\cite{fk2}.
To this end, we first recall the following properties of
the S-matrix for ``physical particles'' \cite{res,zz,fend}
in the overscreening Kondo model. The bulk S-matrix in the spin
sector is decomposed into two parts\cite{res,fend};
the S-matrix for the Gaussian model  and that for the Z$_{n}$ model.
Since the Z$_{n}$ model can be described by
the restricted solid-on-solid (RSOS)
model in the regime I/II\cite{abf,zf},
the corresponding S-matrix is given by the face weight
in the RSOS model \cite{res,ts}.
A remarkable point for the overscreening model is that
the interaction between ``physical particles''
and the impurity is described by the S-matrix
of multi-kinks\cite{fend,mar},
which is given by the fusion of the face
weights of the RSOS model\cite{djm} with the
fusion level $p=n-2S$.
Therefore, nontrivial properties in the
overscreening model are essentially determined by
the RSOS model coupled with the impurity\cite{res,ts}.
Thus the exact FSS can be derived by combining
the spectrum (\ref{eqn:mulfss1}) with that of the RSOS model
coupled with the impurity.
The latter spectrum can be obtained analytically
by using the functional equation method\cite{res2,kp}, in which
the analyticity properties of the transfer matrices
play a central role.
By some technical reasons, we restrict our analysis to the case
of $p=1$ here.  After some manipulations,
we end up with  the FSS of the RSOS model coupled
with the impurity\cite{fk2},
\begin{equation}
E_{RSOS}=\frac{2\pi v}{L}
\biggl(\frac{j(j+1)}{n+2}-\frac{(m+p)^2}{4n} \biggr)
+\mbox{const},
\label{eqn:fsrsos}
\end{equation}
where $ m=2j\quad(\mbox{mod} \,\, 2)$, and
$j=0, 1/2, 1, ..., n/2$.
It is seen that the spectrum fits in with
Z$_n$ parafermion theory, and only the selection rule for
quantum numbers is changed
by the impurity effect, $m \rightarrow (m+p)$.

In order to compare our results with those of Affleck and
Ludwig\cite{aff1}, let us consider the total finite-size spectrum,
which is given by the sum of
eqs.(\ref{eqn:mulfss1}) and (\ref{eqn:fsrsos}).
It is easily seen that the term  $-(m+p)^2/4n$
in (\ref{eqn:fsrsos}) is  canceled by  $S_z^2/n$ in (\ref{eqn:fss1})
by suitably choosing the quantum number for $S_z$.
This in turn modifies the selection rule for
quantum numbers in SU(2)$_n$ Kac-Moody algebra.
Consequently, we find the total spectrum as
\begin{eqnarray}
E= \frac{2\pi v}{L} & \biggl( & \frac{(Q-n)^2}{4n}
+\frac{\tilde j(\tilde j+1)}{n+2}
+(\mbox{flavor part}) \nonumber \\
&&+n_Q+n_s+n_f \biggr), \label{eqn:fss2}
\end{eqnarray}
with new quantum numbers
$\tilde j=\vert j-p/2\vert$ where $\tilde
j=0, 1/2, 1, ... , n/2$. Here
$Q$ and $j$ are the charge and spin  quantum
numbers for free electrons without the Kondo impurity.
We can say that the effect
due to the Kondo impurity is merely to modify the selection rule for
quantum numbers of spin excitations by
$j \rightarrow \tilde j$,
which indeed results in non-Fermi liquid properties.
This is the essence of the Kac-Moody fusion hypothesis proposed
by Affleck and Ludwig\cite{aff1}.
We think that the above result may be
a microscopic description of the fusion
hypothesis for the multi-channel Kondo model.

\section{Summary}

We have investigated critical properties of the Kondo problem
by using the Bethe ansatz method and boundary CFT.
We have obtained analytically the exact FSS of several models
related to the Kondo problem,
such as the Anderson model, the SU($N$) Anderson model,
the {\it s-d} exchange model,
and the multi-channel Kondo model.
The Kac-Moody fusion hypothesis proposed by Affleck and Ludwig
for the FSS of the Kondo problem has been shown to be
consistent with the exact solution. By applying the finite-size
scaling of boundary CFT, boundary critical phenomena
in the Kondo problem have been investigated.

\acknowledgments{}
S.F. and N.K. were partly supported by a Grant-in-Aid from the Ministry
of Education, Science and Culture, Japan.
The work of S.-K.Y. was supported in part by Grant-in-Aid
for Scientific Research on Priority Area 231 ``Infinite Analysis'',
the Ministry of Education,

\end{multicols}
\end{document}